# Direct and in-situ examination of Li$^+$-transport kinetics in isotope-labelled solid electrolyte interphase

Xiaofei Yu[1,†], Stefany Angarita-Gomez[2,†], Yaobin Xu[1,†], Peiyuan Gao[3], Jun-Gang Wang[1], Xin Zhang[3], Hao Jia[4], Wu Xu[4], Xiaolin Li[4], Yingge Du[3], Zhijie Xu[3], Janet S. Ho[5], Kang Xu[5*], Perla B. Balbuena[2*], Chongmin Wang[1*], Zihua Zhu[1*]

[1] Environmental Molecular Sciences Laboratory, Pacific Northwest National Laboratory, Richland, Washington 99354, USA
[2] Department of Chemical Engineering, Texas A&M University, College Station, TX 77843, USA
[3] Physical and Computational Sciences Directorate, Pacific Northwest National Laboratory, Richland, Washington 99354, USA
[4] Energy and Environmental Directorate, Pacific Northwest National Laboratory, Richland, Washington 99354, USA
[5] Battery Science Branch, Energy Science Division, Army Research Directorate, U.S. DEVCOM Army Research Laboratory, Adelphi, Maryland 20783, USA

[†]These authors contributed equally to this work

*Corresponding authors. Email: zihua.zhu@pnnl.gov; chongmin.wang@pnnl.gov; balbuena@tamu.edu; kang_xu@hotmail.com

-2-

**Abstract**

Here, using unique in-situ liquid secondary ion mass spectroscopy on isotope-labelled solid-electrolyte-interphase (SEI), assisted by cryogenic transmission electron microscopy and constrained ab initio molecular dynamics simulation, for the first time we answer the question regarding Li$^+$ transport mechanism across SEI, and quantitatively determine the Li$^+$-mobility therein. We unequivocally unveil that Li$^+$ transport in SEI follows a mechanism of successive displacement, rather than "direct-hopping". We further reveal, in accordance with the spatial-dependence of SEI structure across the thickness, the apparent Li$^+$ self-diffusivity varies from $6.7\times10^{-19}$ m$^2$/s to $1.0\times10^{-20}$ m$^2$/s, setting a quantitative gauging of ionic transport behavior of SEI layer against the underlining electrode as well as the rate limiting step of battery operation. This direct study on Li$^+$ kinetics in SEI fills part of the decade-long knowledge gap about the most important component in advanced batteries and provides more precise guidelines to the tailoring of interphasial chemistries for future battery chemistries.

As the very first interphase-enabled battery chemistry, lithium-ion batteries (LIBs) revolutionized our life in the past three decades, while their commercial success brought the interphase science into the spotlight. In theory, any battery chemistry operating above 3.5 V requires their electrodes to work beyond the thermodynamic stability limits of their electrolytes, whose irreversible reactions with the electrodes have to be minimized via a self-limiting kinetic protection mechanism often known as "solid-electrolyte-interphase (SEI)" that allows the transport of the working ions but simultaneously insulates the electron-tunneling[1]. For rechargeable chemistries, such an SEI must provide the above "services" for $10^3$ charge/discharge cycles (cycle life) or 5~10 years calendar life with little degradation, hence its chemistry, formation mechanism and how it transports cations constitute the core scientific foundations, upon which the modern rechargeable batteries designed for portable electronic devices, electric vehicles or long duration energy storage applications are built. SEI has been widely referred as "the most important but least understood component" in LIBs[2,3].

Since the birth of LIB, SEI has been intensively investigated, which have led to comprehensive understanding about its chemistry, morphological structure, formation mechanism and degradation, as represented by efforts from Aurbach[4], Besenhard[5], Xu and their coworkers[6]. However, two critical questions regarding interphases remain little understood: (1) How does $Li^+$ move through such an inhomogeneous composite consisting of crystalline, amorphous and polymeric matrices? and (2) How fast can $Li^+$ move across interphase? These questions appear to be particularly puzzling because, from the macroscopic perspective, all LIB can deliver decent to high power density depending on the chemistry and cell configuration, indicating fast $Li^+$-transport across the entire cell, including the interphases, while from microscopic and chemical perspective, those known interphasial components identified so far (fluorides, oxides, carbonates, semi-carbonates, etc.) are rather poor ionic conductors beside being electron insulators.

Regarding the first question, there have been two major hypotheses (Supplementary Fig. 1), neither of which has been directly verified via experiments: (1) "direct-hopping" pathway, in which active $Li^+$ from the electrolyte migrate across the SEI via certain channels, such as grain boundaries in a hetero- and polymicrophasic interphase, without displacing those $Li^+$ already immobilized in the lattice of those non-conducting fluorides, carbonates, semi-carbonates[7]; and (2) "knock-off" pathway, in which $Li^+$ successively replaces those $Li^+$ "immobilized" in the lattice and pushes the "Dominos" to move ahead, which is similar to "Grotthus Mechanism" obeyed by proton traveling in aqueous electrolytes but at much lower rate due to the large size of $Li^+$ [3,8,9].

To the second question, various computational efforts have been made[10-13], placing the $Li^+$-diffusivity within SEI in the order of $10^{-17}$-$10^{-11}$ $m^2/s$, but so far these numbers have been pure speculations without direct experimental verification, due to the fact that SEI is a nanometric entity formed in-situ upon cell activation, which is extremely sensitive to external perturbation and impossible to be studied as a standalone component. While $Li^+$ diffusivity in SEI is apparently high enough to support high-rate cell reactions, its direct measurement requires a characterization technique that is not only in-situ by nature, but also of high temporal and spatial resolutions simultaneously. Most importantly, the technique should be able to differentiate the origin of the $Li^+$, i.e., whether it is from the bulk electrolyte ("active $Li^+$") or from the $Li^+$ thought to be immobilized within SEI during its formation ("passive $Li^+$").

In this work, by applying a newly established in-situ liquid secondary ion mass spectroscopy (SIMS) technique on an SEI labeled with Li isotopes (i.e., $^6Li$ and $^7Li$, Fig. 1), we answer these fundamental questions. Further, we reveal the dynamic natures of SEI, shedding light on how the passive $Li^+$ and active $Li^+$ couple and move through the interphases, and provide fundamental knowledge in guiding the tailoring of interphasial chemistries for future battery chemistries.

## Results and Discussion

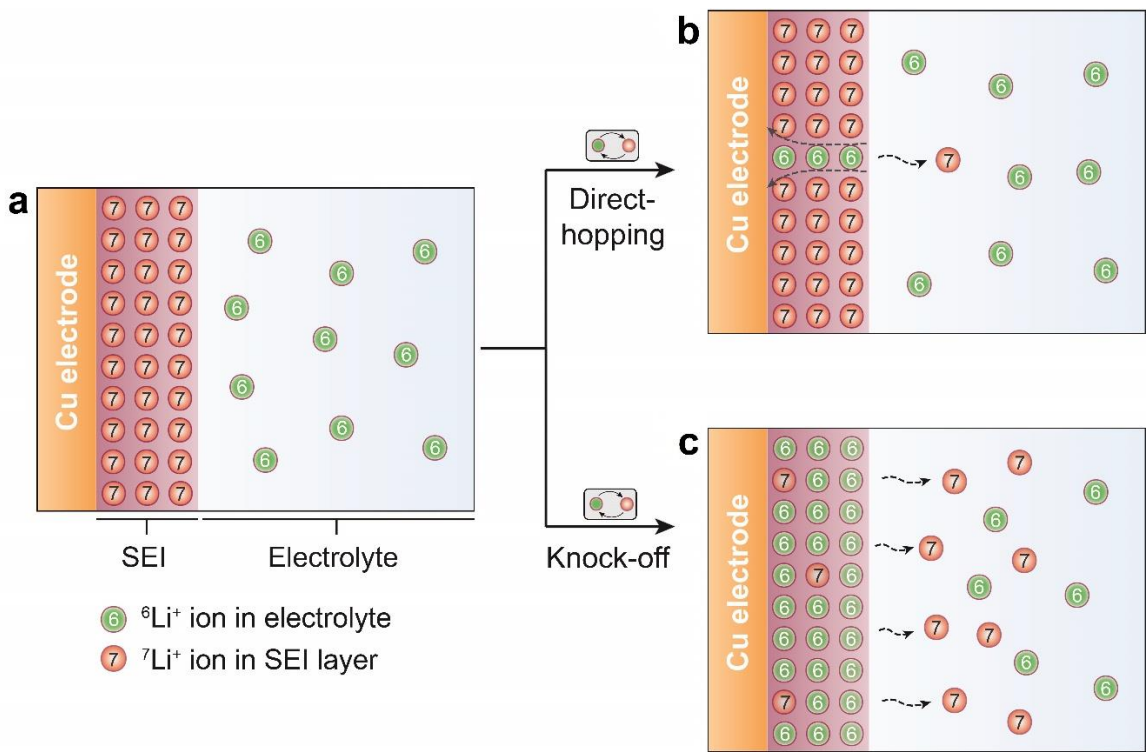

**Fig. 1| A conceptual illustration of differentiating Li$^+$ diffusion mechanisms in SEI with isotope-labeling in SEI and electrolyte. a**, An SEI is formed with $^7$Li$^+$ from natural abundance Li ($^7$Li-dominated) salt, subsequently a $^6$Li-enriched $^6$LiClO$_4$-EC:DMC electrolyte is used to soak the $^7$Li$^+$ SEI to observe the exchange behavior of $^7$Li$^+$ and $^6$Li$^+$. **b**, If "Direct-hopping" pathway is dominant, the $^6$Li$^+$ in electrolyte can only replace a small amount of $^7$Li$^+$ ions in the hopping channels. **c**, If "Knock-off" pathway is dominant, the $^6$Li$^+$ in electrolyte can replace most original $^7$Li$^+$ in bulk components of SEI. A newly developed in-situ liquid SIMS is used to detect the $^6$Li-$^7$Li exchange.

## Differentiation of Two Li$^+$ Transport Mechanisms and Determination of Li Diffusivity in SEI.

First, we grew SEI in an "anode-free" battery configuration[14,15] using an electrolyte with natural abundance of $^7$Li with $^7$Li:$^6$Li = 0.925:0.075. Subsequently, this $^7$Li-dominant SEI was exposed to a $^6$Li-enriched electrolyte, during which $^6$Li$^+$ from the electrolyte should diffuse into SEI and replace those $^7$Li$^+$ originally immobilized in the original $^7$Li-dominant SEI. Direct in-situ monitoring of Li$^+$ and Li$^+$ spatial distribution during the above exchange process would yield Li$^+$ diffusion mechanism. As shown in Fig. 1, if the "Direct-hopping" mechanism applies, most $^7$Li$^+$ in the bulk components of SEI should remain static, while only those $^7$Li$^+$ in the hopping channels should be replaced; On the other hand, most $^7$Li$^+$ in SEI would be easily replaced if the "knock-off" mechanism applies, because these

$^7Li^+$ in theory are not permanently immobilized but still capable of dissociating from the SEI components and equilibrating with the incoming $^6Li^+$ from electrolyte. The self-diffusion rate of those $Li^+$ in SEI could be quantitatively determined based on the change in $^6Li/(^6Li+^7Li)$ ratios with controlled diffusion time.

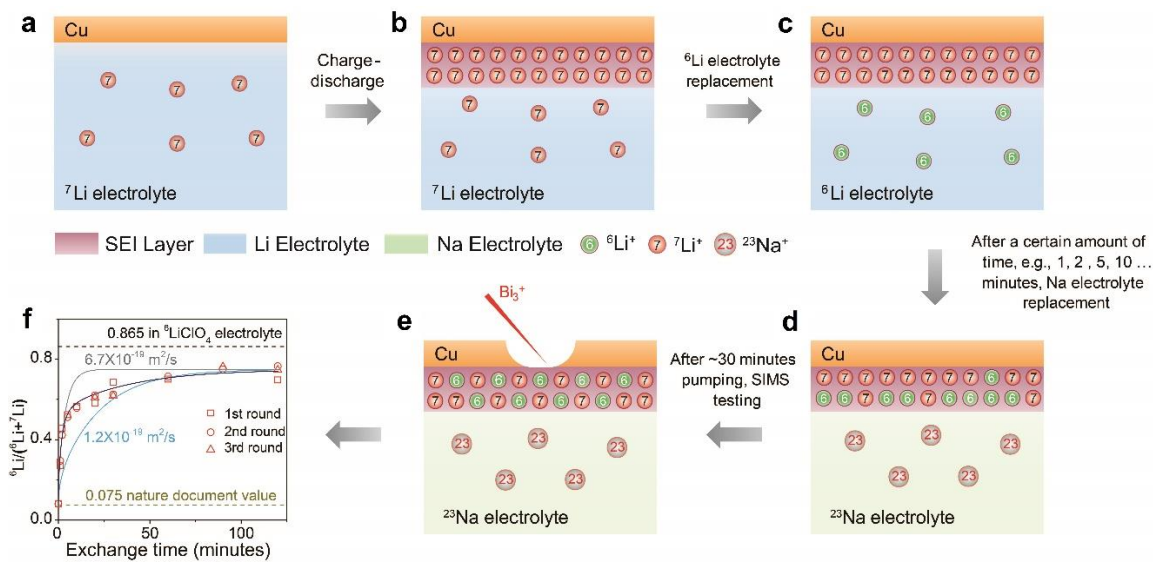

**Fig. 2| Determination of Li$^+$ diffusivity in SEI via in-situ liquid SIMS with Na-electrolyte switch. a**, A Cu anode immersed in a natural abundance electrolyte consisting of $^7LiClO_4$ in EC:DMC. **b**, A $^7Li$-dominant SEI formed on the Cu electrode after three charge-discharge cycles. **c**, The $^7Li$-dominant SEI is exposed to $^6Li$-enriched electrolyte. **d**, After a certain exposure time, such as 1, 2, 5, 10, 20, 30, 60 minutes, a Na electrolyte consisting of NaClO$_4$ in identical solvent mixture was used to replace the $^6LiClO_4$ electrolyte to stop Li$^+$ exchange between SEI and electrolyte. A gradient distribution of $^6Li$ and $^7Li$ in SEI was expected. The NaClO$_4$ electrolyte instead of neat solvent was used here to maintain the same ionic strength so that SEI would not experience additional degradation due to solvent leaching. **e**, The sample was introduced into SIMS instrument for in-situ liquid SIMS analysis. Because self-diffusion of Li$^+$ in SEI does not immediately stop in 30 minutes pumping-down time and such diffusion could be accelerated by extra energy introduced during SIMS measurement, only an average $^6Li/^7Li$ ratio of the whole SEI could be detected in in-situ liquid SIMS analysis. **f**, The relationship between exchange time and measured $^6Li/(^6Li+^7Li)$ ratio values. Three fitting curves are shown, based on 20 nm overall SEI thickness. The grey curve and cyan curve correspond to a singular diffusivity of $6.7\times10^{-19}$ m$^2$/s and $1.2\times10^{-19}$ m$^2$/s across the entire SEI, while the black curve corresponds to a two-layer SEI with distinct diffusivities, in which the outer 2/3 of SEI has a diffusivity of $6.7\times10^{-19}$ m$^2$/s and inner 1/3 of SEI has a diffusivity of $1.0\times10^{-20}$ m$^2$/s.

However, two technical challenges remain here. The primary challenge is that one cannot "turn off" the ion-exchange process in SEI at will; instead, about 30 minutes pumping time is needed before SIMS testing can be performed, during which the ion-exchange still proceeds. Such a delay makes short time exchange testing difficult. The other challenge is that energetic ion beam used in SIMS testing may introduce some extra energy to accelerate $Li^+$ exchange between SEI and electrolyte (Supplementary Fig. 2 and relevant contents in the Supplementary Information). In order to overcome these two challenges and precisely control $Li^+$ exchange time between the $^7Li$-dominant SEI and the $^6Li$-dominant electrolyte, a "chemical switch" is designed, which consists of a non-lithium electrolyte, 1.0 M $NaClO_4$ in the identical salt concentration and solvent composition (Fig. 2), which is based upon the previous discovery that $Na^+$ could not diffuse into a Li-SEI during in-situ liquid SIMS measurement (Supplementary Figs. 2-3), most probably due to the size mismatch between $Li^+$ and $Na^+$[16]. Thus, we expect Na electrolyte to serve as an inert electrolyte that disrupts the $Li^+$ exchange between the electrolyte and the SEI but does not induce additional SEI composition degradation due to the identical ionic strength. Therefore, in the new protocol designed by us, after formation of a $^7Li$-dominant SEI, a $^6Li$-enriched $^6LiClO_4$-EC:DMC electrolyte was introduced into the cell to replace the original $^7Li$-dominant electrolyte, and then after a given interval (e.g., 1 minute or longer), a $NaClO_4$-EC:DMC electrolyte was rapidly pumped in to flush out the $^6Li$-enriched electrolyte. In-situ liquid SIMS conducted immediately thereafter should be able to quantitatively determine the dependence of $^6Li/^7Li$ exchange on SEI-electrolyte exchange time from the $^6Li/(^6Li+^7Li)$ ratio in the SEI. Such a ratio represents the $Li^+$ exchange that has occurred in the fixed duration, and is directly related to the $Li^+$ diffusion rate in the SEI.

Fig. 2f shows the dependence of $^6Li/(^6Li+^7Li)$ ratio in SEI on exchange duration, which was repeated for multiple runs to ensure the reproducibility and confidence in the method, with three runs shown here. An obvious and reproducible time-dependence was established, not only qualitatively verifying that the replacement of $^7Li$ in SEI by $^6Li$ from electrolyte indeed occurs in the expected timeframe as reported previously by Lu et al[9], but also confirming the reliability of the Na electrolyte as an effective chemical switch. The relation takes the shape of a "saturation-like" behavior, with most of the $^7Li$ in SEI (>50%) replaced in the first 5 minutes, followed by a gradually slowing down replacement rate. After 60 minutes, $^6Li/(^6Li+^7Li)$ ratio stabilizes, and stays at about 0.75, indicating that about 84% of $Li^+$ in SEI, although previously believed to be immobilized in the form of various inorganic salts embedded in SEI, are still movable when at equilibrium with $Li^+$ from the electrolyte. Such $Li^+$, although no longer as active lithium reserve in energy storage, should still be considered as "active" for ion transport during the LIB operation. The remaining 16% $Li^+$ ions in SEI are true "dead $Li^+$" and are inert to participate the ion transport during cell reactions[3,17]. The chemical or physical states of these permanently trapped $Li^+$ remain to be explored.

The time-dependence of isotope replacement ratio presents a strong statement regarding how $Li^+$ transport across interphase: it not only indicates that most (>84%) $Li^+$ in SEI layer are replaceable via ion-exchange process, but also reveals how fast (in minute timeframe) this exchange process occurs. This fast kinetics on macroscopic level provide manifestation to the fact that LIB is a chemistry of high rate and power density, and on microscopic level strongly suggests that a "knock-off" mechanism should be the dominant pathway for $Li^+$ transport across SEIs rather than the "direct hopping" mechanism (more details can be seen in the Supplementary information).

**Stratified Structure of SEI**

To quantify $Li^+$ diffusion in SEI, one needs to know the thickness of SEI, which can be determined by SIMS as ~20 nm (Supplementary Fig. 2). However, this thickness is based on an assumption that SEI is sputtered away at the same rate as the reference silicon nitride (SiN), which could deviate from reality. Cryogenic transmission electron microscopy (Cryo-TEM) was used as an external characterization tool to calibrate the morphology and thickness of SEI.

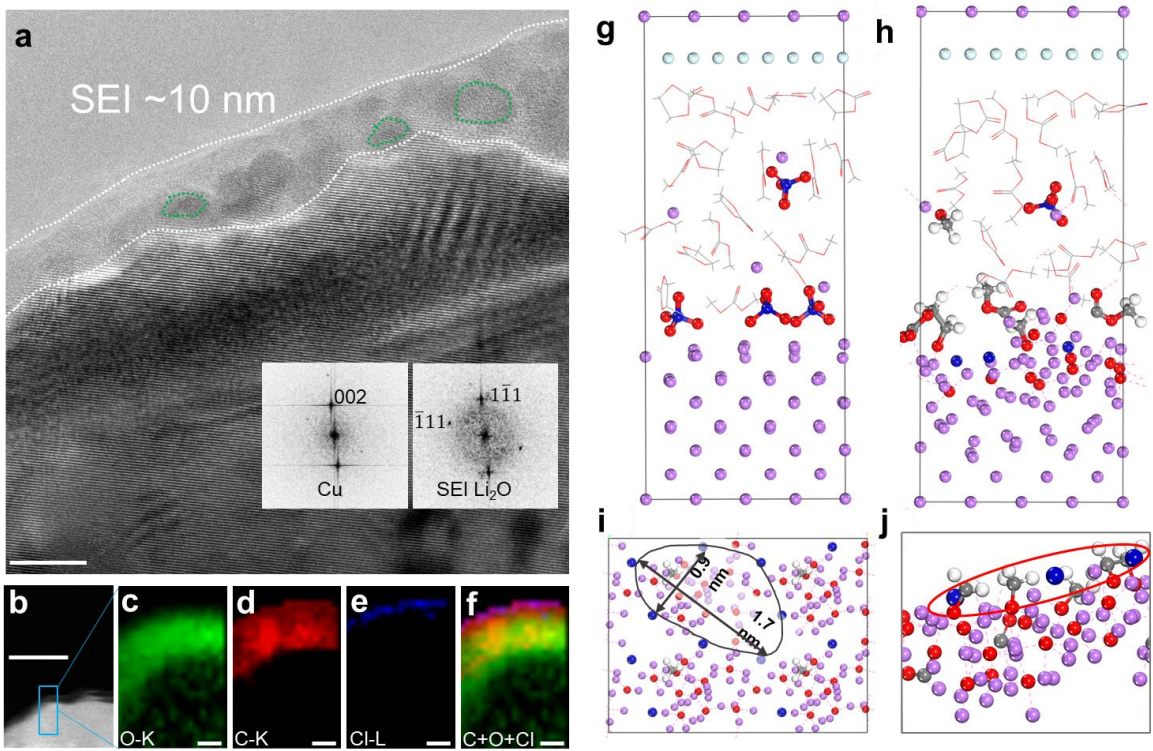

**Fig. 3| Stratified Structure of SEI. a** and **b**, cryo-(S)TEM images, showing the average thickness of the SEI is about 10 nm (average value from multi-locations) and some $Li_2O$ nanoparticles are observed in amorphous matrix. **c-f**, EELS maps show that the SEI can be divided into two layers: the inner SEI is more carbon-depleted, while the outer SEI is more carbon-enriched. Also, chlorine is only observed at the most outside of SEI. **g**, Snapshot of *ab initio* molecular dynamics (AIMD) of SEI formation on Li metal surface in $LiClO_4$-EC:DMC electrolyte at initial state. **h**, Snapshot of AIMD of SEI formation on Li metal surface in $LiClO_4$-EC:DMC electrolyte after formation of SEI AIMD at 4000 fs. **i**, Top view of the cell (2x2) amorphous nanoparticles. **j**, Side view picturing Cl anion on top of the forming SEI. Scale bars, 10 nm in **a**, 50 nm in **b**, and 2 nm in **c-f**.

The cryo-TEM images collected on various spots (Fig. 3) reveal an average SEI thickness of 10 nm, which differs from the SIMS data (Supplementary Fig. 2). The discrepancy appears to be significant, which essentially reflects the different states of SEI that SIMS and cryo-TEM detect respectively. Recently, it has been found that the SEI layer thickness captured by "dry state" cryo-TEM shows a typical thickness of 10 nm, while the SEI thickness in "vitrified state" detected by cryo-TEM is 20 nm due to swelling by a 1.0 M organic carbonate-based electrolyte[18]. The SEI thickness obtained by cryo-TEM in this work represents the "dry state" SEI, while the SEI thickness captured by in-situ liquid SIMS of 20 nm is similar to the "vitrified state" as the SEI is still in contact with the native liquid electrolyte and hence maintains its native morphology as it does in real battery environment. Therefore, a thickness of 20 nm is used in subsequent quantifications in this work.

With the SEI layer thickness of 20 nm and application of Fick's diffusion laws, the in-situ liquid SIMS data can be fitted to derive the Li diffusivity. However, we find that a single diffusivity across the whole thickness of SEI layer cannot faithfully fit all the data points (Fig. 2f, fitting details in the Supplementary information), indicating that the Li$^+$ diffusivity across SEI layer essentially exhibits a graded value, which appears to be consistent with the stratified structure of SEI that has been proposed by numerous work[9,19]. The scanning TEM (STEM)- electron energy loss spectroscopy (EELS) maps in this work (Fig. 3) provide further support for such a stratified SEI structure, featuring an outer layer of ~2 nm with enrichment of Cl, while the inner layer is enriched with C and O (Fig. 3). It should be emphasized here that such chemical distribution highly depends on the electrolyte composition, and would differ in state-of-the-art electrolytes based on fluorinated anions.[3] In this work, Cl only resides at the outmost layer, while the inner of SEI is almost Cl-free, which is consistent with our previous in-situ liquid SIMS observation that anions in the electrolyte and their fragments were hardly observed in the inner impermeable SEI, because their presence in electrolyte-electrode interfacial regions would be repulsed by the negatively charged electrode prior to electrolyte decomposition[15]. Cryo-STEM-EELS images (Fig. 3) reveal that the impermeable inner SEI sublayer can be further divided into two sub-layers: a C-depleted and hence more inorganic sub-layer of 1/3 of the overall SEI thickness which is spatially adjacent to Cu electrode; and a C-enriched and hence more organic sub-layer that resides at the electrolyte side with a thickness 2/3 of the overall SEI. Such a picture enriches the "two-layer" SEI model developed by Qi and Harris[8,20] with new details. As the Cl-enriched SEI layer is permeable to electrolyte and Li$^+$ diffusivity in electrolyte is relatively fast[21], we focus on the Li$^+$ diffusivity (rate-determining step) in the impermeable inner SEI sublayer.

The stratified structural features of SEI provide us foundation for fitting Li diffusivity data accurately using a two-step diffusion process across the SEI: the Li$^+$ diffusion in the C-enriched outer layer has a higher diffusivity, while that in the C-depleted inner layer has a relatively lower diffusivity. Such an assumption leads to a very satisfactory fitting curve (Fig. 2f), in which the Li$^+$ diffusivity in the C-enriched outer is $6.7 \times 10^{-19}$ m$^2$/s and the Li$^+$ diffusivity in the C-depleted inner sub-layer is $1.0 \times 10^{-20}$ m$^2$/s. Such a result indicates that the Li$^+$ diffusivity in SEI is about 4-6 orders of magnitudes higher than in the diffusivities measured in corresponding bulk materials (more details can be seen in the Supplementary information).

These measured values of Li$^+$ self-diffusivity in SEI are much slower than common expectations in Li ion battery field, because diffusivity of $10^{-19} \sim 10^{-20}$ m$^2$/s may potentially lead to very high resistance at the SEI. Before our work, quite a few ex-situ efforts have been performed to experimentally measure the Li$^+$ transport properties across SEI[9,22], associated with extensive computational calculations[8,11-13,20]. For EC:DMC based electrolyte as used in this work, Li$_2$O, LEDC (Li ethylene dicarbonate) and Li$_2$CO$_3$ have been identified to be the major crystalline component of SEI[8,9]. All the Li diffusivity data available from literature were calculated based on specific crystal models in bulk states, typically at 300 K ranging from $1.6 \times 10^{-16}$ m$^2$/s to $4.0 \times 10^{-16}$ m$^2$/s in Li$_2$O, and from $9.0 \times 10^{-15}$ m$^2$/s to $4.7 \times 10^{-17}$ m$^2$/s in Li$_2$CO$_3$[11,13]. It has been even predicted that Li$^+$ diffusivity in Li$_2$CO$_3$ at 300 K can reach $1-5 \times 10^{-11}$ m$^2$/s[12], which is essentially comparable to Li$^+$ diffusivity in liquid electrolytes, typically $10^{-9}$-$10^{-11}$ m$^2$/s. It is apparent that the Li$^+$ diffusivity in SEI experimentally determined in this work is at least three orders of magnitude smaller than these predicted based on previous computational modeling. Despite the widely scattered data in literature, we believe that the directly measured Li$^+$ diffusivity value in this work should be of higher confidence, because we notice that it is comparable to the Li$^+$ diffusivity in LiFePO$_4$, which has been experimentally determined to be ranged from $1.8 \times 10^{-18}$ m$^2$/s to $2.2 \times 10^{-20}$ m$^2$/s at 293 K, depending on lithiation extent as represented by the x number in Li$_{1-x}$FePO$_4$[23-26]. Since LiFePO$_4$ has been generally considered a cathode material of high rate capability in LIB industry[27], its low Li$^+$ diffusivity obviously never constitutes a problem to in the operation of LIB even at high drain rates. Considering that the thickness of SEI is about only 20 nm or less, such a Li$^+$ diffusivity value in SEI is reasonable, while in broader context, it makes much more sense that the Li$^+$ diffusivity in SEI, which is a solid composite by nature, is closer to that in a solid cathode material rather than in liquid electrolytes. Therefore, the low diffusivity of Li$^+$ in SEI often makes SEI the most resistive components in the cell, hence the future effort at emerging battery chemistries should divert more resources in resolving this barrier.

**Li⁺ Transport Mechanism Revealed by Computer Simulation**

Many previous computational works attempted to explain Li$^+$ diffusion behavior in SEI based on crystal structure theories[8,11-13,20], because SEI chemistries as observed via traditional ex-situ tools reveal the existence of micro or nano size crystal particles in the SEI[1-3]. However, in the last 10 years, in situ and cryo-TEM have been used to characterize the SEI, further revealing that the nanocrystals are only the minor part of SEI, while the major part is fully amorphous[28-30]. More interestingly, Zhang et al. observed the "swelling" behavior of SEI[18], which cannot be explained by any traditional crystal/nanocrystal theories. Therefore, assumptions based on traditional crystal theories, such as vacancies and interstitial, may deviate from reality of SEI significantly. In this work, we tried to circumvent the potential problems caused by these concepts and focused on Li$^+$ diffusivity values, which, together with the newly developed computer simulation, leads to understanding of SEI structure on a molecular level.

To gain molecular level insight on Li$^+$ transport mechanism in SEI, we simulate SEI development on Li metal and subsequently Li$^+$ transport kinetics in the SEI layer (Figs. 3g-j, and Fig. 4). SEI formation on Li metal shows three features: Firstly, upon formation of SEI layer, the tiny amorphous nanoparticles emerge in SEI with a size less than 2 nm, which do not grow with time. Secondly, The Cl atoms are rejected from the amorphous clusters, leading to a configuration that Cl is spatially located at the boundaries of the amorphous nanoparticle and encapsulation of amorphous nanoparticle by Cl (Figs. 3i-j). Thirdly, the Cl atoms remain at the top of the SEI layer, which is consistent with cryo-TEM observation, where Cl is spatially segregated at the surface of SEI layer.

The diffusion mechanism of the Li$^+$ in the SEI is revealed by using thermodynamic integration within constrained-AIMD (c-AIMD) simulations in the Blue Moon ensemble[31-33]. Two possible pathways that the Li$^+$ could take as determined from c-AIMD are shown in Fig. 4a. In the first pathway, the Li$^+$ moves around the border of the SEI structure, thus limiting the number of changes induced in the oxide coordination shell. Initially, the Li$^+$ joins the first encountered oxide coordination shell by knocking off other Li$^+$ and displacing them out of the coordination shell (Fig. 4b image 1 and 4), which is consistent with the experimental determination of displacement model. The largest energy barrier (~

0.73 eV) is given by the diffusion of the $Li^+$ along with its first coordination shell, while the $Li^+$ displaces additional $Li^+$ initially in the coordination shell. At the end, the $Li^+$ coordinates with a CO moiety that exists as a product of the electrolyte reduction (Fig. 4b, image 5). In the second pathway, the $Li^+$ moves through the thickest part of the amorphous SEI. In contrast with the previous pathway, the $Li^+$ changes oxide coordination shell with low energy barriers (ranging from 0.09 eV to 0.27 eV) and it coordinates simultaneously with several oxygen atoms from different oxide shells as shown in Fig. 4c (image 4). This mechanism allows a lower-barrier migration of the $Li^+$ to the SEI surface, where similar to the previous case it coordinates with a CO moiety. However, it is important to highlight that many atoms are involved in the diffusion of the $Li^+$ during the second pathway, which could potentially slow down the diffusion process due to the high connectivity of the $Li^+$ with different oxide coordination sites. Complete details of the diffusion mechanisms and corresponding energy barriers can be found in the Supplementary information (Supplementary Figs. 4-5).

$Li^+$ diffusivities derived from each of the pathways are listed in Supplementary Tables 1-2 in the Supplementary information, indicating an overall faster $Li^+$ diffusion on pathway 2 than on pathway 1, with a typical value in the range of $10^{-15}$ to $10^{-19}$ m$^2$/s. Giving the fact that the limiting step for $Li^+$ migration will be the slowest diffusion path, therefore the predicted diffusivity is consistent with the results determined by the in-situ liquid SIMS measurement in Fig. 2f.

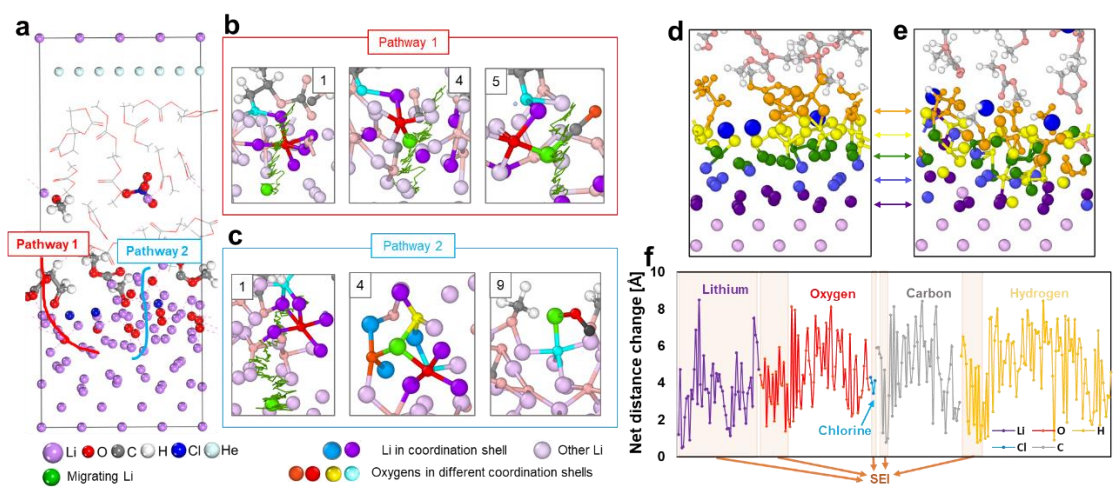

**Fig. 4| Atomic dynamics in SEI. a**, Illustration of the two different pathways of diffusion. **b**, Selected snapshots of the coordination shells found by the migrating lithium in pathway

1. **c**, Selected snapshots of the coordination shells found by the migrating lithium in pathway 2. Images numbers correspond to each pathway description shown in Supplementary Figs. 4-5. **d**, Atom mobility in SEI during the thermodynamic integration process performed for pathway 1 at initial state. **e**, Atom mobility in SEI during the thermodynamic integration process performed for pathway 1 after Li$^+$ movement in pathway 1. **f**, Net distance change for Li, O, Cl, C, and H in SEI (light pink background) and electrolyte (white background). Colors in **d** and **e** showcase a visual comparison of the atom's displacement. Total number of atoms on the SEI: Li=50, C=8, O=20, H=13, and Cl=3.

**Live SEI**

Accompanying the Li$^+$ migration in SEI is the simultaneous relaxation movement of other atoms, typically such as O, C and H. For example, the average migration distance of O atoms is about 3.5 Å (Fig. 4f). Such an observation indicates that the O atoms can "travel" in SEI, as the Li-O distances in Li$_2$O and Li$_2$CO$_3$ are about 1.9-2.1 Å[34,35]. Such deviation of individual atoms from their lattice position has been typically observed in certain solid state electrolytes via a so-called "paddle wheel mechanism", where the rotational or vibrational movement of coordination sites assist the working ions to travel with lower energy barrier. These thermal relaxational movement of O in amorphous phase is expected to occur at much higher amplitude than their inorganic lattices. Migration of O is in markedly contrast with the case of most inorganic oxide materials, in which the anions have relatively large sizes, and they normally behave as immobile matrix, especially compared with the singly charged small Li$^+$. Similar case appears to be true for most solid electrolyte materials for LIBs with highly mobile Li$^+$ in low mobile anion matrix[36]. Apparently, the active movement of O/C/H in SEI suggests that the SEI is highly dynamic ever than we have been able to capture with temporal and spatial resolution under in-situ condition.

The high dynamic behavior of most atoms in SEI leads to a situation that there are few ordered structures observed in SEI, and only amorphous nanoparticles with a size of around 1-2 nm can be observed (Fig. 3i). Such a result is well consistent with our cryo-TEM observations, in which the main body of SEI is amorphous. More importantly, under such a situation, most (80+%) Li$^+$ (as well as C/O/H) are existing on particle surface, and they can be replaced relatively easily, because it has been well known that atomic/ionic diffusivity at nanoparticle surface can be several orders of magnitude higher than dense bulk materials[37].

A critical question is why compact and ordered structures cannot form in SEI. In our computer simulation, various organic moieties in SEI, including two CO moieties (charge ~ -3|e|), a $CO_2$ (~ -3|e|), a $C_2O_2H_4$ (~ -2|e|), and three $COH_3$ (~-1|e|), are found. Such organic moieties are decomposition products of solvent molecules. It is well known that organic reactions are normally complex with many by-products and difficult to form certain stoichiometric ratio compounds with long range order. A possible explanation is that the diversity of such organic moieties makes SEI a high-entropy system with dynamic nature, and unfavorite for growth of any compact and ordered structure. Therefore, if more components can be introduced into SEI to further raise the entropy, e.g., introducing more F atoms into SEI[38-40], the performance of SEI as an ion conductor might be improved.

## Conclusions

Despite the important role of SEI in advanced batteries, its key properties in transporting working ions have never been directly measured with any reliable characterization tools, while most reported $Li^+$ diffusivity in SEIs remains speculative based on crystalline models and ignoring the heterogeneous nature of SEI. In this work, we designed an in-situ study to directly measure $Li^+$ diffusivity across SEI and identified it to be in the range of $6.7\times10^{-19}$ $m^2/s$ and $1.0\times10^{-20}$ $m^2/s$ at room temperature. This value is significantly lower than what have been predicted by previous MD simulations and quantum chemistry calculations, but within comparable range with that of an active electrode material. Further, against what has been assumed that other species in SEI remain stationary as a matrix while the active $Li^+$ are migrating for the battery function, it is apparent that with $Li^+$ migration, other coordinating species in the SEI exhibited certain localized mobility and cooperate with $Li^+$ transport, thus rendering $Li^+$ motion with less energetic barrier. The new knowledge represents a significant step forward in our understanding of SEI and will provide more precise guideline to the efforts of designing a better SEI.

## Methods

**Cell preparation**

The battery cell was fabricated on a polyether ether ketone (PEEK) block using the method reported in our previous publication[15]. In brief, a liquid chamber with a size of 6.0 mm (L) × 5.5 mm (W) × 1.0 mm (H) was machined on the PEEK block with two liquid channels for introduction of electrolytes. The Li ion battery cathode, a $LiCoO_2$ layer (~55 µm thick) coated on a thin Al foil (~15 µm thick), was immobilized at the bottom of the liquid chamber. The anode, a ~70 nm thick Cu film, was sputter-coated on a 100 nm thick $Si_3N_4$ membrane, which was immobilized on a silicon frame of 7.5 mm (L) × 7.5 mm (W) × 0.2 mm (H). The silicon frame (with $Si_3N_4$ membrane and Cu anode below it) was placed on top of the liquid chamber and it was sealed using an epoxy glue. The effective cathode area was about 10.0 $mm^2$ and the effective anode area was about 4.0 $mm^2$, and the distance between them was about 0.8 mm. Two thin Cu wires were attached with the cathode and anode, respectively, using Ag paste so the battery could be charged-discharged. Following the assembling of the cell, a desirable electrolyte could be introduced into the liquid chamber in an argon-filled glove box. After being sealed, the Li ion battery cell could be loaded on a ToF-SIMS sample holder (as shown in Supplementary Fig. 6), and then introduced into the SIMS instrument for operando analysis.

**Formation of initial SEI with natural-abundance Li isotopes**

A 1.0 M $LiClO_4$ (with natural-abundance Li isotopes) in 1:2 (v/v) ethylene carbonate:dimethyl carbonate (EC:DMC)) electrolyte was introduced into the battery cell. A constant-current mode ($1.5 \times 10^{-6}$ A) was used for charging-discharging in this work. A Li ion battery cell could be charged-discharged for at least 10 cycles. A typical charge-discharge curve is shown in Supplementary Fig. 7. The voltage quickly (less than 40 s) increased to 1.7 V, and then gradually increased to ~3.7 V (~1300 s) until it slightly decreased to reach a relative stable value at 3.6 V. After further charging of about 800 s, the battery was discharged to 0.5 V to make it sure that most Li metal was stripped from the Cu anode. Three charging/discharging cycles were performed to make it sure a good SEI layer formed on the Cu electrode surface.

**$^6$Li replacement in SEI**

After the 3$^{rd}$ discharge (all active Li metal was stripped), the natural-abundance 1.0 M LiClO$_4$ in EC:DMC electrolyte was replaced by a 1.0 M $^6$LiClO$_4$ in EC:DMC electrolyte. In this research, SIMS data showed that the $^6$Li/($^6$Li+$^7$Li) was 0.86±0.01 in the $^6$LiClO$_4$ electrolyte, while $^6$Li/($^6$Li+$^7$Li) was 0.078±0.004 (document value 0.075) in the natural abundance LiClO$_4$ electrolyte. The SEI layer was thin and Li$^+$ diffusion from the electrolyte into the SEI layer was quick, while about 30 minutes pumping time was needed before in-situ liquid SIMS measurement. Therefore, after a desirable time (1, 2, 5, 10, 20, 30, 45, 60, 90, 120 minutes in this research) of introduction of $^6$LiClO$_4$ electrolyte, a 1.0 M NaClO$_4$ in EC:DMC electrolyte was used to replace the $^6$LiClO$_4$ electrolyte, so Li$^+$ ion exchange between the electrolyte and the SEI could be stopped. Then the battery cell was sealed and loaded onto a sample holder for in-situ liquid SIMS analysis.

### *In-situ* liquid SIMS

The in-situ liquid SIMS measurement was same as we described in our previous paper[15,41]. In brief, a pulsed 25 kV Bi$_3^+$ beam with the beam size of ~450 nm in diameter was used for all measurement. The pulse frequency was 10 kHz, the pulse width was about 150 ns, and the corresponding beam current was ~0.36 pA. The incident angle of the primary Bi$_3^+$ beam was 45 degrees off the normal. For each measurement, the Bi$_3^+$ beam was scanned on a round area of ~2 μm in diameter around the center of the Si$_3$N$_4$ membrane to drill a hole on it. The measurement was stopped when stable liquid signals were observed (normally 50-200 s after liquid signals were detected). A mass spectrum, depth profiles, and 2-D ion images were simultaneously collected, and 3-D ion maps were available after reconstruction of raw data. The pressure in the analysis chamber was about $5 \times 10^{-7}$ to $1 \times 10^{-6}$ mbar with sample in. During SIMS measurements, the pressure change normally was negligible. A low energy electron flood gun (~10 eV, 1 μA) was used to control any potential charging during measurement. It should be noted that only unit mass resolution spectra could be obtained in this work[41]; however, no interreference signals existed around $^6$Li$^+$ and $^7$Li$^+$, so such a low mass resolution was acceptable. Because Li$^+$ signals could be very strong, some detector saturation might occur, while such saturation could be corrected during data analysis. To ensure the data reproducibility, at least two cells for each time point were tested. The isotopic ratio results were reproducible for most time points, as shown in Fig. 3 in the main text.

**Cryo-TEM details**

Cu foil for TEM observation was directly prepared from commercial Cu foil which was used as the current collector in coin cell. It was thinned by a Gatan precision ion polishing system (PIPS, Gatan, USA) with Ar ion milling to make an electron transparent thin area. After preparation, the Cu TEM foil was transferred into an Ar-filled glovebox to avoid oxidation and assembled in a CR2032 coin cell. A polyethylene separator was used to separate the Cu and $LiCoO_2$ electrodes. The same electrolyte and electrochemical conditions with SIMS were used for SEI formation (Arbin BT-2000). After SEI formation, the Cu TEM foil was taken out of the coin cell and slightly rinsed with DMC to remove trace electrolyte in the glovebox. Then, the Cu TEM foil with formed SEI was placed in a sealed bag fulfilled with Ar. The sealed bag was plunged directly into a bath of liquid nitrogen after taken from the Ar-filled glovebox until the Cu TEM foil reach to very low temperature (around 100 k). We then quickly took out the Cu TEM foil sample from the sealed bag and loaded onto a pre-cooled Gatan cryo-holder (Elsa, Gatan, USA) using a cryo-transfer station to ensure entire process occurred under cryogenic environment. TEM observations were performed on a 300 kV FEI Titan monochromated (scanning) transmission electron microscope ((S)TEM) equipped with a probe aberration corrector. The samples were viewed at low temperature (100 K) under low dose condition (~1 $e·Å^{-2}·s^{-1}$ for low magnification imaging, and ~1000 $e·Å^{-2}·s^{-1}$ for high resolution TEM imaging). Multi-locations were imaged, showing the average thickness of the SEI was about 10 nm (as shown in Fig. 3 and Supplementary Fig. 8). Spectroscopy experiments were performed on a Gatan GIF-Quantum spectrometer. The EELS collection semiangle during the spectroscopy experiments was ~45 mrad. EELS spectra dispersion was 0.25 eV/channel with vertical binning at 130 in dual EELS mode. The probe beam current was around 25 pA, and pixel dwell time was 0.001-0.2 s.

**Computer simulation details**

The simulation cell measures 10.3 Å x 13.8 Å x 33.3 Å and it is formed by seven layers of lithium metal in (100) facet where the bottom two layers are fixed to resemble bulk behavior. A Helium layer is added at the top of the cell to prevent interaction due to periodic boundary conditions. The electrolyte is formed by EC and DMC in a 1:2 molar ratio and 2.54 M of $LiClO_4$ placed relatively close to the metal slab to promote the formation of the SEI.

Calculations were performed with Vienna ab Initio Simulation Package (VASP)[42-44]. In this study different types of calculations including initial optimizations, Ab Initio Molecular Dynamics (AIMD), and Thermodynamic integration[31-33] implemented in the Bluemoon ensemble was performed. the projector augmented wave (PAW)[45,46] was used to describe the electron-ion interactions and the Perdew-Burke-Ernzerhof generalized gradient approximation (GGA-PBE)[47] was used for exchange-correlation functional. For optimizations and AIMD, a k-point mesh used for the surface Brillouin zone integration was Monkhorst-Pack[48] of 2x2x1 and 1x1x1 for the thermodynamic integration calculations. An NVT canonical ensemble at 298K with a time step of 1 fs for a total of 4000 fs was used for AIMD simulations. The Nose thermostat[49,50] with a damping parameter of 0.5 was used. The energy cur off for the plane-wave basis was 400 eV and Gaussian smearing width of 0.05 eV. For the thermodynamic integration calculation, the reaction coordinates also called collective variable ($\xi$) is the motion of the $Li^+$ ion from an initial location ($\xi_1$ inside the Li metal slab) towards a defined location ($\xi_2$ top of the SEI structure at the interphase with the electrolyte) with a small step size of 0.0008 Å every femtosecond. Every step in this pathway provides a free energy gradient ($\delta F/\Delta \xi$), the value of the free energy gradient is obtained by averaging 100 fs, and the free energy $\Delta F$ is calculated as an integral along the path.

The diffusivity values shown in Supplementary Tables 1-2 are calculated using the Diffusion pre-factor ($D_0$) for lower temperature diffusion in $Li_2O$[15] and the energy barriers found in this work.

It should be noted that a higher molar concentration (2.54 M $LiClO_4$ instead of 1.0 M) was used in the simulations to minimize the computational expenses. A more diluted electrolyte requires a much larger simulation cell. However, using a lower salt concentration would not affect change the reported conclusions since the results shown here mainly depend on the Li transport through the amorphous Li-oxide phase. The simulations show that other SEI components appear distributed as shown in the experiment.

**Acknowledgements:** This work was supported by Laboratory Directed Research and Development (LDRD) programs (majorly an FY 16 Open Call LDRD and partially Chemical Dynamics Initiative) at Pacific Northwest National Laboratory (PNNL). CMW and WX thank the support from the Assistant Secretary for Energy Efficiency and Renewable Energy, Vehicle Technologies Office (VTO) of the U.S. Department of Energy (DOE) under the Advanced Battery Materials Research (BMR) Program and the US-Germany Cooperation on Energy Storage with the Contract No. DE-AC05-76RL01830. KX and JH thank the funding from JCESR, an Energy Hub funded by DOE Office of Science. PBB and SA-G thank the support from the Assistant Secretary for Energy Efficiency and Renewable Energy, VTO of the U.S. DOE through the BMR Program (Battery500 Consortium Phase 2) under Contract No. DE-AC05-76RL01830 from PNNL. The research was performed on a project (Award DOI: 10.46936/staf.proj.2016.49687/60006123) from the Environmental Molecular Sciences Laboratory (EMSL), a national scientific user facility sponsored by the DOE's Office of Biological and Environmental Research and located at PNNL. PNNL is operated by Battelle for the U.S. DOE under the Contract DE-AC05- 76RL01830. Computational resources from the Texas A&M University High Performance Research Computing are gratefully acknowledged. The authors appreciate beneficial discussion with Yue Qi of Brown University.

**Author Contributions:** Z.Z., C.M.W. and K.X. conceived the project. X.Y. prepared the battery cells and conducted the in-situ liquid SIMS characterizations under instruction of Z.Z. X.Y. and J.W. organized SIMS data and drew relevant figures. S.A.G. performed computer simulation under instruction of P.B.B. Y.X. performed cryo-TEM measurement under instruction of C.M.W. P.G. and Z.X. fitted in-situ SIMS data to quantify Li$^+$ diffusivity. J.H. and K.X. synthesized and purified the isotope-labelled lithium salts. W.X., H.J. and X.L. provided the relevant electrolytes. Z.Z., X.Y., S.A.G. and Y.X. drafted the